\documentclass[12pt,draftclsnofoot,onecolumn]{IEEEtran}

\usepackage{amssymb}
\usepackage[cmex10]{amsmath}
\usepackage{stfloats}
\usepackage{graphicx}
\usepackage{subfigure}
\usepackage{tabularx}
\usepackage{epsfig,epsf,color,balance,cite}
\usepackage{verbatim}

\usepackage{algorithm}
\usepackage{algpseudocode}
\usepackage{amsmath}

\usepackage{threeparttable}
\usepackage{diagbox}
\usepackage{booktabs,graphicx}

\usepackage{epstopdf}
\usepackage{epsfig}
\begin{document}
%\title{
%    {\huge Distributed and Streaming Machine Learning for NOMA}}
\title{  { Federated Learning for 6G: Applications, Challenges, and Opportunities}}

\author{
\IEEEauthorblockN{
Zhaohui Yang,
 Mingzhe Chen,
Kai-Kit Wong, \IEEEmembership{Fellow, IEEE},
H. Vincent Poor, \IEEEmembership{Fellow, IEEE}, and Shuguang Cui, \IEEEmembership{Fellow, IEEE}
                  }
\thanks{This work was supported in part by the U.S. National Science Foundation under Grant CCF-1908308.}
\thanks{Z. Yang  is with the Centre for Telecommunications Research, Department of Engineering, King's College London, WC2R 2LS, UK, Email: yang.zhaohui@kcl.ac.uk.}
\thanks{M. Chen  is with the Department of Electrical Engineering, Princeton University, Princeton, NJ, 08544, USA, and also with the Chinese University of Hong Kong, Shenzhen, 518172, China, Email: mingzhec@princeton.edu.}
\thanks{K.-K. Wong is with the Department of Electronic and Electrical Engineering, University College London, WC1E 6BT London, UK, Email: kai-kit.wong@ucl.ac.uk.}
\thanks{H. Vincent Poor is with the Department of Electrical Engineering, Princeton University, Princeton, NJ, 08544, USA, Email: poor@princeton.edu. }
\thanks{S. Cui is with the Shenzhen Research Institute of Big Data and School of Science and Engineering, the Chinese University of Hong Kong, Shenzhen, 518172, China, Email: robert.cui@gmail.com.}
 }

%\author{
%\IEEEauthorblockN{Zhaohui Yang\IEEEauthorrefmark{1},
%                  Mingzhe Chen\IEEEauthorrefmark{2},
%                  Yuanwei Liu\IEEEauthorrefmark{3},
%                  Walid Saad\IEEEauthorrefmark{4}
%                  and Mohammad Shikh-Bahaei\IEEEauthorrefmark{1}
%                  }\\
%                   {
%\IEEEauthorblockA{\IEEEauthorrefmark{1}Centre for Telecommunications Research, Department of Informatics, King's College London, London WC2B 4BG, UK,  email: \{yang.zhaohui, m.sbahaei\}@kcl.ac.uk}\\
%\IEEEauthorblockA{\IEEEauthorrefmark{2}Beijing Key Laboratory of Network System Architecture and Convergence,
%Beijing University of Posts and Telecommunications, Beijing, China 100876, email: chenmingzhe@bupt.edu.cn.}\\
%\IEEEauthorblockA{\IEEEauthorrefmark{3}School of Electronic Engineering and Computer Science, Queen Mary University of London, London E1 4NS, UK,  email: yuanwei.liu@qmul.ac.uk}\\
%\IEEEauthorblockA{\IEEEauthorrefmark{4}Wireless@VT, Bradley Department of Electrical and Computer Engineering, Virginia Tech, Blacksburg, VA, USA,  email: walids@vt.edu.}
% }}
\pagestyle{headings} \maketitle \thispagestyle{empty}

%\newpage
%
%
%\section{Background and Motivation}
\begin{abstract}
Traditional machine learning is centralized in the cloud (data centers). Recently, the security concern and the availability of abundant data and computation resources in wireless networks are pushing the deployment of learning algorithms towards the network edge. This has led to the emergence of a fast growing area, called federated learning (FL), which integrates two originally decoupled areas: wireless communication and machine learning.
In this paper, we provide a comprehensive study on the applications of FL for sixth generation (6G) wireless networks. First, we discuss the key requirements in applying FL for wireless communications. Then, we focus on the motivating application of FL for wireless communications.
We identify the main problems, challenges,  and provide a comprehensive treatment of implementing FL techniques for wireless communications.
%In essence, this paper is the first detailed magazine that constitutes a unified reference on  applying distributed learning in NOMA networks.
\end{abstract}

\section{Background and Overview on Federated Learning for Wireless Communications }

\subsection{Motivation}

%1. First machine learning, then distributed optimization, next federated learning (FL), why FL, how it can be achieved.

%2. Why FL is important for wireless communications.

Due to the explosive growth in data traffic, machine learning and data driven approaches have recently received much attention and are anticipated to be a key enabler for the to be developed sixth generation (6G) wireless networks \cite{saad2019vision}.
Nowadays, standard machine learning approaches require centralizing the training data on a single data center or cloud.
Since massive data samples need to be uploaded to the data center,  transmission delay can be very high and user privacy is not guaranteed in standard centralized machine learning approaches.
However, low-latency and privacy requirements are important in the emerging application scenarios, such as unmanned aerial vehicles,  extended reality (XR) services, % (including augmented, mixed, and virtual reality (AR/MR/VR)),
autonomous driving, %, and remote surgery,
which makes   centralized machine learning approaches inapplicable.
Moreover, due to limited communication resources, it is impractical for all the wireless devices
that are engaged in learning to transmit all of their collected
data to a data center   that uses
a centralized learning algorithm for data analytic or network
self-organization.

Therefore, it becomes increasingly attractive to process data locally at edge devices.
This has led to the emergency of distributed optimization methods.
In distributed optimization, each node can compute on its own data and sends the results to its neighbours or a central node.
Distributed optimization has many applications, such as user selection optimization, resource allocation optimization, trajectory optimization, and distributed machine learning design\cite{chen2019joint}.

\begin{figure}[t]
\centering
\includegraphics[width=5in]{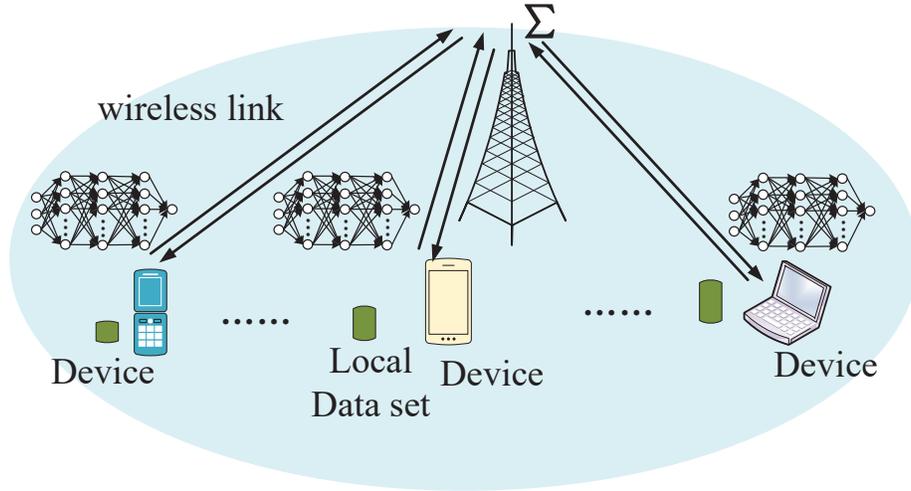}
%\vspace{-2em}
\caption{A FL algorithm over wireless communication systems.} \label{fig1-0}
%\vspace{-2em}
\end{figure}

Combining the advantages of distributed optimization and machine learning,  distributed learning frameworks are needed to enable wireless devices to collaboratively build a shared learning model with training taken place locally. One of the most promising distributed learning algorithms is the
emerging \emph{federated learning} (FL) \cite{konevcny2016federated,huang2020federated,zhu2019broadband,zhu2020one,zeng2020energy,amiri2020machine,gunduz2020communicate,
amiri2020federated,hosseinalipour2020federated,hosseinalipour2020multi,jin2020design,liu2020privacy,kassab2020federated,kairouz2019advances,
samarakoon2019distributed} framework is anticipated in future Internet of Things
(IoT) systems.
 In FL, wireless devices can cooperatively execute a learning task by
only uploading local learning models to the base station (BS) instead of sharing the entirety of
their training data, as illustrated in Fig. \ref{fig1-0} \cite{yang2019eeFL}.
Since the data center cannot access the local data sets at the users, FL can protect data privacy of the users.

For wireless communications, FL has the following advantages:
(i)  exchanging local machine learning model parameters instead of the massive training data saves energy and consumes less wireless resources;
(ii) training machine learning model parameters locally can effectively reduce transmission latency;
(iii) FL preserves data privacy since  the training data remains at each device and only the local machine learning model parameters are uploaded;
(iv) using different learning processes to train several classifiers from distributed data sets increases the possibility of achieving higher accuracy especially on a large-size domain;
(v) FL is inherently scalable since the growing amount of data may be offset by increasing the number of computers or processors, and providing a natural solution for large-scale learning where complexity and memory are the main obstacles.

FL can be used to solve complex convex and nonconvex optimization problems that arise in various use cases such as network control, user clustering, resource management, and interference alignment. Besides, FL enables users to collaboratively learn a shared prediction model while keeping their collected data on their devices for user behaviour predictions, user identifications, and wireless environment analysis.
Based on the predicted results, the BS can efficiently allocate the wireless resources for the devices.

\subsection{Classification}

For FL, there are two main classifications: federated reinforcement learning (FRL) and
federated supervised learning (FSL).
In \cite{liu2019lifelong}, the goal of FRL is to enable wireless devices to remember what they have learned and what
other wireless devices have learned.
FRL can be used in the case where multiple wireless devices make decisions in different environments.
In FRL, each wireless device builds a learning network with the help of other wireless devices.
\begin{enumerate}
\item[1.] Initially, one edge device first obtains its private strategy model learning network through reinforcement learning (RL) in its own environment and then uploads it to the BS  as the shared model.
\item[2.]  After a while, %wireless devices desire to learn by RL in new environments.
the wireless devices download the shared model from the BS as the initial actor model in RL.
Wireless devices get their own
private learning networks through RL
in new environments.
After
training is completed, wireless devices upload their private learning networks  to the BS.
\item[3.]  At the BS,
the private learning networks are fused into the shared model, and then a new shared model will be generated.
The new shared model can be used by other wireless devices. Other
wireless devices will also upload their private learning networks to the BS to evolve and update the shared model.
\end{enumerate}

The FSL technique builds a uniform learning model through iteratively updating information between the BS and wireless devices, where the local private data is fully labeled.
The FSL procedure
contains three steps at each iteration: local computation
at each wireless device, local FSL model parameters transmission from each wireless device, and
result aggregation and broadcast at the BS.

\begin{enumerate}
\item[1.] Every wireless device needs to compute the result by using its fully labeled  data set locally.
\item[2.] All wireless devices upload the local prediction parameters to the BS via wireless links in the uplink.
\item[3.]  The BS aggregates the prediction model parameters and
broadcasts the global prediction model parameters to all the wireless devices in the downlink.
\end{enumerate}

\subsection{Relevant Surveys and Our Contributions}

%Related surveys, contributions and organizations.

There are some interesting surveys  about FL in wireless communications such as \cite{li2020federated,lim2020federated,murshed2019machine,wang2019edge,park2019wireless,aledhari2020federated}.
The unique characteristics and challenges of FL were discussed in \cite{li2020federated}.
Moreover, this work  provided an overview of the current approaches, and outlined several directions of future work.
The work in \cite{lim2020federated} introduced the challenges of FL implementation and reviewed the
existing solutions.
In \cite{murshed2019machine}, the authors
 described the challenges of machine learning systems that are configured at the edge computer networks.
Considering  RL,
the authors in \cite{wang2019edge}
proposed to integrate deep RL techniques and the FL framework with mobile edge systems, for optimizing mobile edge computing, caching and wireless communication resource.
In addition, the work in \cite{park2019wireless} explored the key building blocks of edge machine learning and different wireless network architectural splits for wireless communications.
The study about FL application was surveyed in \cite{aledhari2020federated} including software and hardware platforms, protocols, real-life applications and use-cases.

\begin{table}[t]
\centering
\caption{An overview of selected surveys about FL in wireless communications.} \label{tab:complexity}
\begin{tabular}{ccc}
  \hline
  \hline
  Subject  &   Contributions&   Related Work \\ \hline
  FL & Introductory tutorial on unique characteristics and challenges of FL&    \cite{li2020federated}\\
  FL& Challenges of FL implementation  &    \cite{lim2020federated}\\
  Edge machine learning& Challenges of machine learning systems at the edge computer networks &    \cite{murshed2019machine}\\
  FL&  FL and RL for optimizing mobile edge computing and caching &    \cite{wang2019edge}\\
  Edge machine learning&  Edge machine learning architectures  &    \cite{park2019wireless}\\
  FL&  FL application and use-cases &    \cite{aledhari2020federated}\\
  \hline
  \hline
\end{tabular}
\end{table}

We aim to gather the state-of-the-art contributions that address the key challenges of applying
FL techniques for wireless networks. In particular,
our objectives are three-fold:  to provide a comprehensive descriptions of FL algorithm,
  to identify the key open problems in wireless communication that can
be addressed using FL methods, and  to point out the emerging applications in wireless communication with FL.

\section{Performance and Requirements for Federated Learning}

%\subsection{Federated Learning Architecture for Wireless Communications}

\subsection{Performance Evaluation}
 \begin{figure}[t]
\centering
\includegraphics[width=6in]{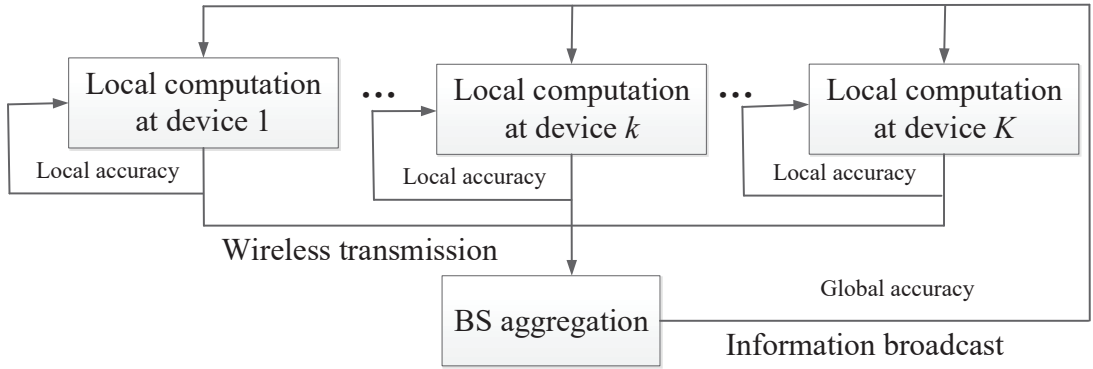}
%\vspace{-2em}
\caption{FL procedures over wireless networks.} \label{sys2}
%\vspace{-2em}
\end{figure}
The procedure of FL over wireless networks is shown in Fig.~\ref{sys2}.
The FL procedure contains three steps at each  iteration: local computation at each device (using several local iterations), local FL parameter transmission for each device, and result aggregation and broadcast at the BS.
The local computation step is essentially the phase during which each device calculates its local FL parameters by using its local data set and the received global FL parameters.
There are four main performance indicators for FL: delay, energy, reliability, and massive connectivity.

\subsubsection{Delay}
\begin{figure}[t]
\centering
\includegraphics[width=5in]{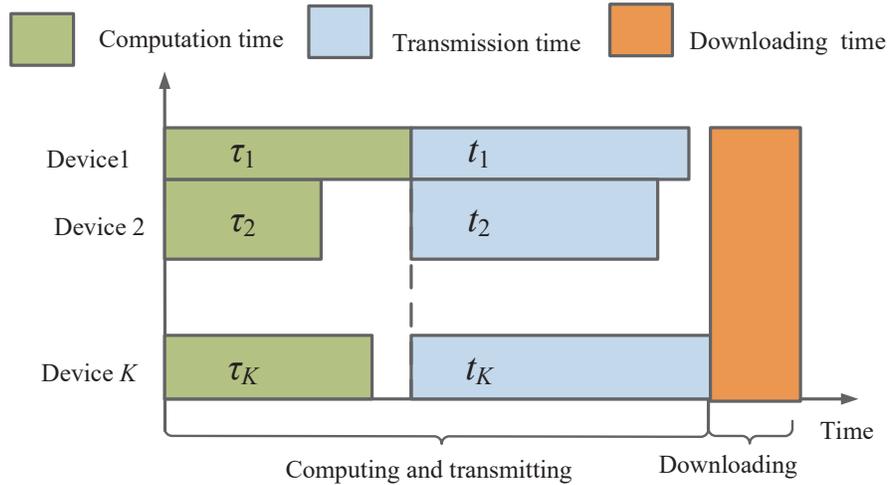}
%\vspace{-2em}
\caption{Energy performance of FL over wireless networks.} \label{sys2-2}
%\vspace{-2em}
\end{figure}
%{\color{myc1}one figure (4 subfigures) for FL example: a) components of delay, b) components of energy, c) short frame structure for reliability, d) random scheduling policy}
According to Fig.~\ref{sys2-2}, the delay of FL includes: local computation delay of wireless devices, uplink transmission delay, BS aggregation delay, and downlink transmission delay.
%(1) delay (see the metric in URLLC)
Considering the tradeoff between
local computation delay and wireless transmission delay, it is of importance to minimize the delay for FL via joint transmission and computation optimization.

\subsubsection{Energy}
Due to limited energy budget of wireless devices, both
local computation energy and transmission energy must be considered during the FL process.
The calculation of local computation energy involves the number of iterations for local computation at each device, and the transmission energy is related to the number of iterations for the FL algorithm to converge.

\subsubsection{Reliability}
To train an FL algorithm in a distributed manner, the
devices must transmit the training parameters over wireless links which can introduce training
errors, due to   limited wireless resources (e.g., bandwidth) and the inherent unreliability of
wireless links.
%Since all training
%parameters are transmitted over wireless links, the quality of the training will be affected by wireless
%factors such as packet errors and the availability of wireless resources.
 % considering Gaussian errors in wireless communication
For example, symbol errors introduced by
the unreliable nature of the wireless channel and by resource limitations can impact the quality
and correctness of the FL updates among users. Such errors will, in turn, affect the performance
of FL algorithms, as well as their convergence speed.

\subsubsection{Massive connectivity}
To meet the low latency requirement of FL,
we need to collect data distributed among a huge number
of devices rapidly through wireless communications.
However, with enormous number of devices, conventional
interference-avoiding channel access schemes become infeasible
since they normally result in excessive latency.
To overcome
this challenge,
over-the-air computation is a
promising approach for fast wireless data aggregation via exploiting
the superposition property in a multiple access channel \cite{8952884}.
\subsection{Potential to Meet 6G Requirements}
It is expected that 6G communication systems will hence to accommodate
  125 billion   wireless devices
by 2030.
As a result, it is important to develop an
automatic data processing
framework to allow   edge learning to take place.
As one of the key enabling technologies, FL has the potential to meet the following 6G requirement  \cite{saad2019vision}. %: massive ultra-reliable, low latency communications (mURLLC), scalable architecture, and human-centric services.

\subsubsection{Massive ultra-reliable, low latency communications (mURLLC)}
Due to the explosive growth in the number of wireless devices in 6G, the 5G URLLC requirements will be changed to the mURLLC.
With FL, multiple edge computing units can be used to cooperatively learn a shared model for the network, which can decrease service delay and provide high reliability.

\subsubsection{Scalable architecture}
Different from a central cloud, %which is built based on a architecture architecture
edge intelligence, such as FL, is built in a distributed manner, which includes many edge servers with computing and communication capabilities.
To serve a massive number of devices in the future 6G,
it is important to provide a scalable and
decomposable architecture to allow
simultaneous computing among
multiple edge servers.
It is expected that the FL architecture
will play an important role in the future 6G services and applications.

\subsubsection{Human-centric services}
Different from the rate-reliability-latency metrics in 5G, 6G involves human-centric services, which requires quality of experience related to the physical movement of the users.
FL can be used to predict the movements and gestures of users, and the BS can utilize the predicted results to improve the quality of experience for users.

\section{Federated Learning for Wireless Communications: Motivating Applications}

%Due to ultra dense deployment of small cells in future networks, inter-cell interference usually leads to nonconvex resource allocation problems.
Machine learning tools can exploit big data analytic for wireless network state estimation and find the relationship between the optimized variables and objective functions in an online manner so as to reduce the computational complexity for solving the nonconvex problems in wireless communication.
Besides, machine learning is powerful because it can optimize problems that no one knows how to describe the problems.
However, given that multi-cell network needs global channel state information (CSI),
centralized learning algorithms may require the BSs to continuously upload their collected data to a centralized processing server, which can lead to a high network overhead and significant delays.
As a consequence, using a centralized learning algorithm for resource management or network control may need a large number of iterations to converge.
Thus, centralized machine learning algorithms will not be able to handle the resource allocation, signal detection and user behaviour prediction problems in future networks.
To this end, FL is needed, which enables users or BSs to manage the resource in a distributed manner and analyze their collected data locally.

\subsection{ Driving Application of FL for Wireless Problems}

%%%=============================================0802
%{\color{myc1}{Why FL, the advantages, the problems that can be solved via FL}}
\subsubsection{Resource management}
Spectral efficiency and connectivity optimization of multi-cell network  always leads to nonconvex resource allocation problems, which were often solved by conventional algorithms such as successive convex approximation and matching theory with high complexity and impractical implementation.
Therefore, there is a need to introduce new FL techniques that can be used to address a variety of resource management challenges such as distributed power control for multi-cell networks, joint user association and beamforming design, and dynamic user clustering.

\begin{figure}[t]
\centering
\includegraphics[width=5in]{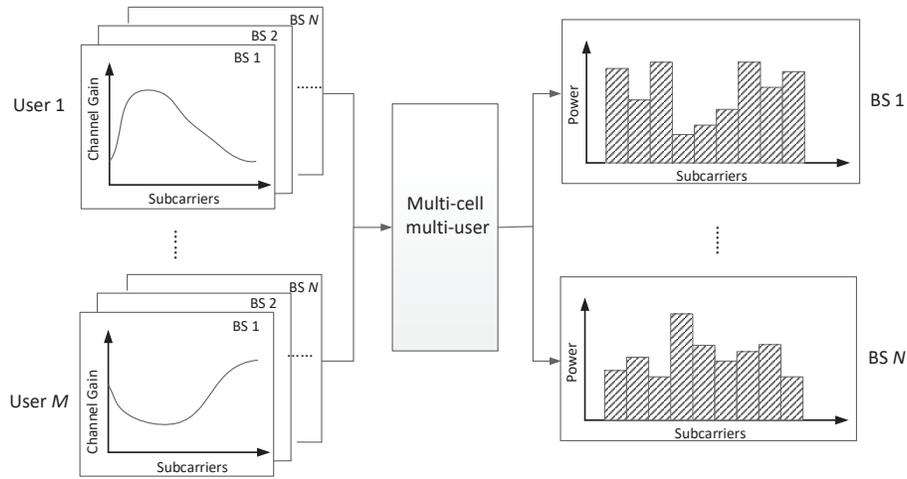}
%\vspace{-2em}
\caption{Multi-cell power control problem.} \label{fig-mcpower}
%\vspace{-2em}
\end{figure}

For multi-cell power control, as shown in Fig. \ref{fig-mcpower},
FRL enables each BS to build the relationship between the power control schemes and utility values so as to find the optimal power control scheme.
In FRL, the BSs on a connected network process data locally by minimizing small optimization problems, and exchange the local results among the neighbors to arrive at a global solution.

%
% connectivity -  dynamic clustering
% user clustering supervised learning
% Then, we investigate how distributed subgradient descent based machine learning algorithms  can be used to study the user association and beamforming of a multi-antenna network.
Further, FRL can be used for dynamic user clustering, where users individually learn the clustering parameters by RL and the BS builds the unified clustering parameters based on the received clustering parameters from all users.
%Finally, we use an example to show, in details, how distributed RL can be implemented for bandwidth and power allocation as well as user association.

\subsubsection{ User behavior predictions}
Due to the heterogeneous quality-of-service requirement of users, user behaviour prediction is of great importance for the implementation of wireless networks.

FL can be used to predict the users behaviors such as mobility patterns where each user performs a local FL algorithm to train the learning model using its own user behavior data and upload the trained model to the BS. Then the BS generates and broadcasts the unified FL model parameters to all users. Based on the mobility predictions, the users can dynamically choose a subchannel to upload data in the uplink, the BS dynamically allocates multiple subchannels to multiple users in the downlink, and multiple users which occupy the same subchannel can perform non-orthogonal multiple access (NOMA) or full duplex.

The quality of service of users can be predicted by FL, where each BS uses the FL algorithm based on its stored information such as users' requested data, gender, job, and device type and all BSs transmit the  FL model results to a server to get a unified FL model.

\subsubsection{Channel estimation and signal detection}

Channel estimation and signal detection is a major challenge due to the random features of wireless channels in wireless communication networks.
% additional intra-cell interference %, decentralized and large-scale nature
%of NOMA networks.
%Broadly speaking, signal detection consists of multi-user detection.
For downlink systems, FL algorithms are used for
 channel estimation and multi-user detection, where each user performs a FL algorithm for channel estimation and signal detection, and sends
their local FL model parameters to the BS that will generate the global
FL model.
For multi-cell uplink systems, multi-user signals can be  detected  via iteratively transmitting individually FL model parameters from all BSs to a server and broadcasting the unified FL model parameters from the server to all the BSs.
Further, FL algorithms  can be utilized to
%parallelizing support vector machine (SVM) algorithm, where multiple users perform parallel computation and transmit results to the BS,
automatically design the codebook of BSs and decoding strategy of users to minimize the bit error rate, where users upload the learned result to the corresponding BSs and the BSs forward their unified learned result to a server.

\subsection{Reconfigurable Intelligent Surface}

\begin{figure}[t]
\centering
\includegraphics[width=5in]{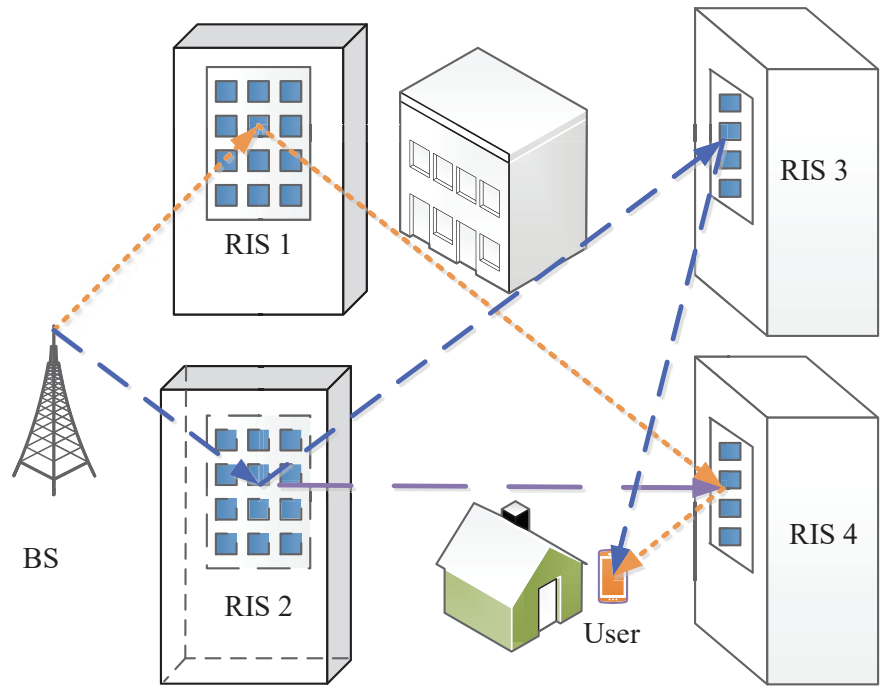}
%\vspace{-2em}
\caption{A RIS-assisted wireless communication system.} \label{fig1}
%\vspace{-2em}
\end{figure}

Reconfigurable intelligent surface (RIS)-assisted wireless communication has been proposed as a potential solution for enhancing the energy efficiency of wireless networks \cite{basar2019wireless,zhang2019capacity,8580675,pan2019intelligent2,nadeem2019large,wei2020joint,8741198,chongwenDL2020,
huang2019holographic,yu2019robust,zheng2020double,chaccour2020risk}. %qingqing2019towards
An RIS is a meta-surface equipped with low-cost and passive elements that can be programmed to turn the wireless channel into a partially deterministic space.
In RIS-assisted wireless communication networks, a  BS sends control signals to an RIS controller so as to optimize the properties of incident waves and improve the communication quality of users.
The RIS acts as a reflector and does not perform any digitalization operation.
Hence, if properly deployed, an RIS promises much lower energy consumption than traditional amplify-and-forward (AF) relays~\cite{hum2013reconfigurable,huang2014relay,ntontin2019reconfigurable}.
%Since no amplifier is used in the RIS, the
However,
the constraint on the diagonal phase shift matrix and unit modulus of the reflecting RIS makes the
joint design of transmit beamforming and phase shifts extremely challenging.
To address high-dimension, complex EM environment, and mathematically intractable nonlinear
issues of communication systems, the model-free FL method as an extraordinarily
remarkable technology can be used.

\subsubsection{CSI Detection}

In the RIS-enhanced system, to achieve the full advantages
of the architecture, several efficient technologies are required
including the joint active and passive beamforming, resource
allocation, and energy-efficient design. It is noted that all of
above designs rely on the perfect CSI between the BS and RIS, and the perfect CSI between the RIS and users.
 However, it
is infeasible for the RIS-enhanced systems to estimate the
accurate CSI when the radio frequency (RF) chains or sensors
are not equipped on the RIS.  To this end, it is meaningful to use FL for CSI detection in RIS-assisted wireless communications.

The FL-based model training approach can be used in RIS-assisted massive multiple-input-multiple-output (MIMO) systems
 \cite{elbir2020federated}.
 The FL approach mainly includes three steps:  data collection, training and prediction.
 In the first step, each user collects its local training data set, where the pilot sequence is the input and the received signal is the output.
 Then, each user computes the updated model by using its own local data set, and the BS generates a global model after receiving the updated models from all users.
 In the last step,
each user estimates its own channel by feeding the
received pilot data into the trained model.

\subsubsection{Distributed joint passive and active beamforming}
In RIS, the phase shift of each RIS element can be adjusted to improve the performance of RIS-assisted wireless communication systems.
Different from conventional communications, it is of importance to jointly optimize the passive beamforming (phase shift matrices at the RIS) and active beamforming (beamforming at the multi-antenna transmitter)
\cite{yang2020federated,ni2020federated}.
To solve the complicated joint passive and active beamforming, deep learning (DL) has been used to design the best reflection matrix
of RIS elements in indoor communication environments \cite{huang2019indoor}.
In practice, similar to multi-hop relaying systems, multiple
RISs can be used to overcome severe signal blockage between the BS and users to achieve
better service coverage.
The authors in \cite{huang2020hybrid} presented a multi-hop RIS-assisted communication scheme to overcome the
severe propagation attenuations and improve the coverage range at Terahertz (THz) band frequencies, where
the hybrid design of transmit beamforming at the BS and phase shift matrices is obtained by
the advances of RL.
Due to the high complexity of using centralized RL, FRL can be utilized to solve the joint passive and active beamforming problem, where all users can individually  optimize the phase shift matrices and transmit beamforming via RL, and the BS broadcasts the aggregated learning model to all users.

\subsubsection{Phase shift prediction}
Due to the randomness of wireless communication channels, it is required to adjust the phase shift matrixes as the wireless channel changes.
By exploiting the time-correlated  property of channel fading, the phase shift matrixes of the RIS can be predicted with FL.
To predict the phase shift, each user uses long short-term memory (LSTM) network to predict the future CSI and phase shift matrices using local data set, while the BS aggregates the received results from all users.

%\subsection{Extended Reality}

\subsection{Semantic Communication}
Semantic communication, is similar to a brain communication,
where the difference between meaning of the transmitted
symbols and that of recovered ones is correlated \cite{xie2020deep}.
This correlation can be useful for joint encoding and decoding when the bandwidth of the system is limited or the bit error rate is high for some typical communication systems.

\subsubsection{Channel encoder and decoder design}
Using semantic communication technique 
which enables the devices only to transmit semantic information to the server, 
rather than traditional bit or symbol, the network bandwidth utility can be effectively improved.
However, semantic communication model requires the training data from multiple distributed devices, which induces huge communication cost for data transmission. To solve this problem, a FL based DL enabled  semantic communication can be proposed for channel encoder and decoder design.
First, a DL  model can be used to extract the semantic information from text or audio with robustness to noise.
Then, in an FL approach, the devices and the server obtain practicable DL models with the server aggregating devices locally trained models and sending the aggregated model back to the devices.

\subsubsection{Distributed semantic communication for IoT}

The emerging technologies, such as smart city,  IoT and machine to machine (M2M)
networks, require the intelligent communication between different ends, such as human to machine.
For those applications, the intelligent communication depends on the background and interface language model \cite{xie2020lite}.
Besides, there are always a large number of devices in IoT.
The above factors motivate the design of distributed semantic communication for IoT with FL.
The distributed semantic communication with FL includes three steps.
In the first step, the BS computes the semantic communication model using DL.
In the second step, the BS broadcasts the trained DL model to all users.
In the third step, each user obtains the semantic features through receiving the broadcast information. Then, each user transmits the semantic features to the BS and the BS accordingly updates the semantic communication model.

\subsection{XR}
\begin{figure}[t]
\centering
\includegraphics[width=4in]{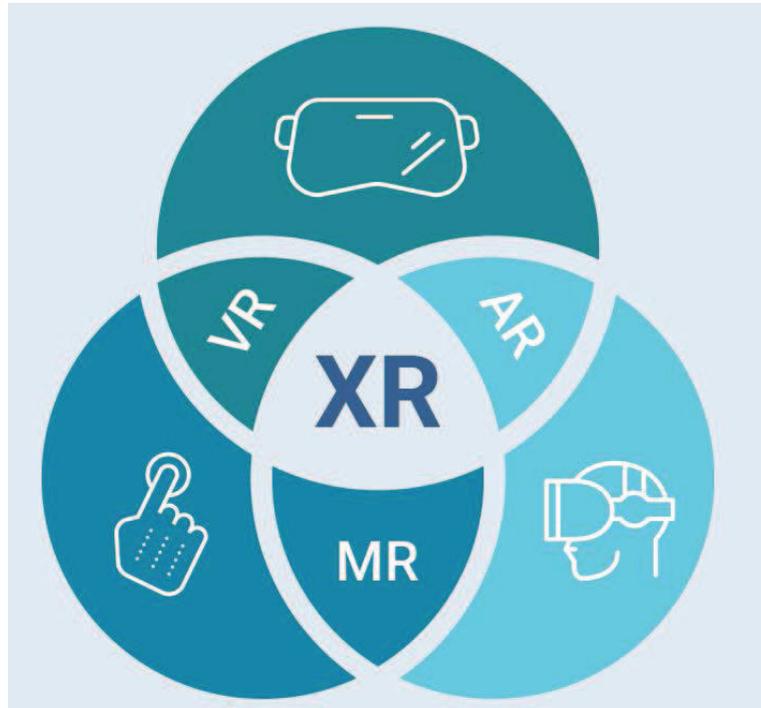}
%\vspace{-2em}
\caption{Classification of XR.} \label{fig-xr}
%\vspace{-2em}
\end{figure}

XR refers to all real-and-virtual environments generated by computer graphics, which includes augmented reality (AR), mixed reality (MR), and virtual reality (VR), as in Fig. \ref{fig-xr}.
Deploying XR over wireless communication networks is an essential step for realising XR applications \cite{saad2019vision}.
Due to the seamless and immersible requirements, it is important to introduce wireless communication technologies to meet the stringent quality of service requirements, such as high data rate and ultra low latency.
For XR allocation over wireless communications, the location and orientation information needs to be sent to the BSs and the BSs construct the 360 degrees images for users based on the received information

\subsubsection{User movement prediction}
 In a wireless XR network, the user body movement can heavily influence the wireless resource allocation and network management\cite{chen2019federated}.
 To deal with the user movement challenge, FL is effective at predicting the users' movements and actions.
Based on the predicted movements and actions, the BSs can improve the generation
of the XR images and optimize the resource management for
wireless XR users.

\subsubsection{Resource allocation}
FL can be used to develop self-organizing algorithms for solving dynamic resource
management problems for XR networks \cite{8755300}. In particular,
FL can be used to adaptively
optimize the wireless resource and construct the  format of the XR images based on the wireless environment.
\subsection{Non-Orthogonal Multiple Access}

%1. Interference management in scalable multi-cell NOMA

%2. Joint user association, beamforming and decoding order design for NOMA

%3. Dynamic user clustering and resource allocation in NOMA based mobile networks

%4. Robust design for NOMA and rate splitting multiple access (RSMA) networks

NOMA is envisioned to be a promising technique for the development of next-generation wireless networks \cite{Zhiguo2015Mag}.
By serving multiple users at the same time and frequency resource,
NOMA can scale up the number of served users,
increase spectral efficiency, and improve user-fairness compared to existing orthogonal multiple access (OMA) techniques.
Recently, significant research efforts have appeared focusing on various challenge of NOMA~\cite{Liu2017Proceeding,Liu2018MIMONOMA,Qin2018Mag}, that include modeling, performance analysis, signal processing, and emerging NOMA applications such as heterogenous networks (HetNets), cognitive radio networks and millimeter wave (mmWave) communications.
%As NOMA brings an additional power or code domain, %on top of the exiting resource domain,
%it brings sophisticated %intra/inter cell/cluster
%interferences to the existing networks \cite{Dai2015NOMA}.
The non-orthogonal resource allocation nature of
%the interference issue is particularly challenging for multi-antenna NOMA design \cite{Liu2018MIMONOMA}, multi-cell NOMA networks \cite{Shin2017Mag}, hybrid NOMA networks, etc.
%More particularly, the  heterogenous nature of
NOMA  necessitates the introduction of novel models and algorithms for addressing several challenges that include:
joint user clustering and resource allocation for devising a scalable multi-cell NOMA design, advanced channel estimation and signal detection for large-scale NOMA networks, and dynamic user behaviour prediction in NOMA based mobile networks.
%1) Interference mitigation and  advanced multi-user detection (MUD)  for devising a scalable multi-cell NOMA design \cite{Shin2017Mag}; 2) Joint user allocation and beamforming design for multi-antenna NOMA to provide superior performance compared to conventional OMA networks \cite{Liu2018MIMONOMA}; and 3) Dynamic user clustering and resource allocation in NOMA based mobile networks.

%admm, distributed convex optimization with machine learning
% dnn,
% nn
% cnn
%
%
%

%heterogenous quality of service
%In order to address the aforementioned challenges, advanced statistical and intelligent tools as provided by
%machine learning \cite{chen2017machine,andrieu2003introduction,freeman2000learning,collobert2008unified,bishop2006pattern} will play a major role in shaping the design of NOMA.
%Due to the heterogeneous requirements of NOMA, NOMA brings the additional complexity compared to OMA networks, which usually leads to nonconvex problems
%
%....the machine learning...
%Due to non-orthogonal resource allocation and heterogeneous requirements of NOMA, NOMA brings the additional complexity compared to OMA networks, which usually leads to nonconvex problems.
Due to non-orthogonal resource allocation, intra-cell interference always exists in NOMA networks compared to OMA networks, which usually leads to nonconvex resource allocation problems.
Traditional optimization methods, which are used to solve the nonconvex problems for optimizing the performance of NOMA networks, mostly operate in an offline manner with high computation complexity and depend largely on accurate CSI \cite{yang2017on,Yang2018Power,Shin2017Mag,ni2020resource}.
 Machine learning tools \cite{andrieu2003introduction,freeman2000learning,collobert2008unified,bishop2006pattern} can exploit big data analytics for wireless network state estimation and find the relationship between the optimized variables and objective functions in an online manner so as to reduce the computational complexity for solving the nonconvex problems in NOMA.
  %
%advanced statistical and intelligent tools as provided by machine learning \cite{chen2017machine,andrieu2003introduction,freeman2000learning,collobert2008unified,bishop2006pattern} will play a major role.
%To address the  in NOMA, machine learning tools  are undoubtedly one of the most important tools, as evidenced by the wide adoption of machine learning tools in a myriad of applications domains \cite{chen2017machine,andrieu2003introduction,freeman2000learning,collobert2008unified,bishop2006pattern}.
%Compared to traditional optimization method used to optimize the performance of NOMA, which operates in an offline manner with high computation complexity and depends largely on accurate channel state information (CSI) \cite{yang2017on,Yang2018Power,Shin2017Mag},
%machine learning will be able to operate in a fully online manner by learning.
%machine learning is advantageous in providing a data-driven approach to efficiently obtain the optimal solution, detecting sharply changing CSI and exploiting adaptive learning features.
However, given that multi-cell NOMA needs global CSI,
%%%%%%%%%%%%%%%%%%%%%%%%%%%%%%%%%%%%%%%%%%%%%%0327
centralized learning algorithm may require the BSs to continuously upload their collected data to a centralized processing server, which can lead to a high network overhead and significant delays.
%Moreover, uploading all of their collected data to a centralized processing server will result in significant delays. %, which can seriously degrade the superior performance of NOMA in delay.
%give a reason why NOMA cannot tolerate large delay).
%Besides, machine  learning  models over vast amount data  in  one  machine  are  impossible, especially for the case that the data  is distributed  across  different  locations.
%The data privacy among different BSs also contributes to the impossibility of the centralized learning algorithm.
%{\color{blue}(Think about other reasons in NOMA why centralized learning cannot work).}
%Besides, due to huge number of users in uplink grant-free NOMA, the joint user activity and user detection in one machine requires extremely high computational resource and memory.
Besides, in NOMA, each subcarrier can be occupied by multiple users. In consequence, using a centralized learning algorithm for resource management or network control may need a large number of iterations to converge.
Thus, centralized machine learning algorithms such as in \cite{7001098,8219722,chen2016caching,8372711} will not be able to handle the resource allocation, signal detection and user behaviour prediction problems in NOMA.
%{\color{blue} ( add NOMA challenges from 1) to 4) in the first paragraph.)}
%To this end, a distributed learning framework is needed, which enables users or BSs to manage the resource in a distributed manner and analyze their collected data locally.
%To this end, a distributed learning framework that can be trained by the collected data at each BS and cooperatively build a learning model for resource management, user detection, user classification and clustering, as well as CSI prediction is needed.
%Compared to centralized learning algorithms, distributed learning algorithms have three advantages:
%\begin{itemize}
%\item using different learning processes to train several classifiers from distributed data sets increases the possibility of achieving higher accuracy especially on a large-size domain;
%\item learning in a distributed manner provides a natural solution for large-scale learning where algorithm complexity and memory limitation are always the main obstacles;
%\item distributed learning is inherently scalable since the growing amount of data may be offset by increasing the number of computers or processors.
%\end{itemize}
For NOMA, FL have two important use cases: 1) FRL can be used to solve complex convex and nonconvex optimization problems that arise in various NOMA use cases such as network control, user clustering, resource management and interference alignment %{\color{blue}( add some applications that are specific for NOMA)},
and 2) FSL enables users to collaboratively learn a shared prediction model while remaining their collected data on their devices
 for user detection %, user association, user clustering,
  and CSI prediction. %{\color{blue}( add some applications that are specific for NOMA)}.

\subsubsection{Resource management in NOMA}
%\item\textbf{Distributed Learning for Resource Management in NOMA (5 pages)}
%\begin{itemize}
%\item Interference management in scalable multi-cell
%NOMA  with distributed learning.
%\item Joint user association, beamforming and decoding order design for NOMA with distributed learning.
%%\item User decoding order for NOMA with distributed learning.
%\item Dynamic user clustering and resource allocation in NOMA based mobile networks with distributed learning.
%\end{itemize}
%admm
%distributed subgradient descent algorithm   using the parameter server :Communication Efficient Distributed Machine Learning with the Parameter Server \cite{li2013parameter}

%By splitting users in the power or code domain, NOMA can be used to simultaneously serve multiple users at the same frequency or time resource.
With superposition coding at the transmitter and successive interference cancellation (SIC) at the receiver,
NOMA can achieve higher spectral efficiency than OMA.
Moreover, %high connectivity is another feature since
NOMA can serve multiple users
at the same resource (e.g., time/frequency) by exploiting the
user differences in the power domain \cite{Yang2017Energy,Yang2018EEIoT}.
This power domain feature provides
rich opportunities for NOMA to support massive connectivity
and meet the users' diverse quality of service.

%%One of the main challenges in NOMA networks is to strike
%%an attractive compromise between the bandwidth efficiency
%%and the energy efficiency of the networks by intelligently
%%controlling the power allocation of the superimposed signals \cite{Liu2017Proceeding}, dynamically
%%scheduling the users for the subchannels or by forming
%%spatially correlated clusters.
%
%% definition, characteristics
%
% spectral efficiency - distributed power allocation
%

The spectral efficiency and connectivity optimization of NOMA  always leads to nonconvex resource allocation problems, which were solved by conventional algorithms such as successive convex approximation and matching theory with high complexity and impractical implementation \cite{Liu2017Proceeding}.
Therefore, there is a need to introduce new distributed learning techniques that can be used to address a variety of resource management challenges such as distributed power control for multi-cell NOMA \cite{Shin2017Mag}, joint user association and beamforming design \cite{Qin2018Mag}, and dynamic user clustering \cite{Cui2018Unspuer}.
For multi-cell power control, FRL enables each BS to build the relationship between the power control schemes and utility values so as to find the optimal power control scheme.
%In the distributed setting, the agents on a connected network process data locally by minimizing small optimization problems, and exchange the local results among the neighbors to arrive at a global minimization solution.
%
% connectivity -  dynamic clustering
% user clustering supervised learning
FRL can also be used to study the user association and beamforming of a multi-antenna NOMA network \cite{li2013parameter}.
Further, the use of FRL for dynamic user clustering in NOMA, where users individually learn the clustering parameters by RL and the BS builds the unified clustering parameters based on the received clustering parameters from all users.

\subsubsection{Channel estimation and signal detection in NOMA}
%\item\textbf{Distributed Learning for Channel Estimation and Signal Detection in NOMA (3 pages)}
%\begin{itemize}
%\item  Channel estimation and multi-user detection for downlink NOMA with distributed learning.
%\item   Multi-user detection in multi-cell uplink NOMA with federated learning. % Error propagation in successful interference cancellation.
%\item   Joint codebook and decoding strategy design for code-domain NOMA networks.
%\end{itemize}
Channel estimation and signal detection in NOMA is a major challenge due to error propagation in SIC for NOMA networks.
% additional intra-cell interference %, decentralized and large-scale nature
%of NOMA networks.
%Broadly speaking, signal detection consists of multi-user detection.
FSL algorithms can be used for
 channel estimation and multi-user detection in downlink NOMA networks, where each user performs a supervised learning (SL) algorithm for channel estimation and signal detection of multiple users due to SIC and send
their local federated learning model parameters to the BS that will generate the global
FL model.
As in \cite{bekkerman2011scaling}, FSL can detect multi-user signal in multi-cell uplink NOMA networks via iteratively transmitting individually learning model parameters from all BSs to a server and broadcasting the unified learning  model parameters from the server to all BSs.
Further, FSL can be used to
%parallelizing support vector machine (SVM) algorithm, where multiple users perform parallel computation and transmit results to the BS,
automatically design the codebook of BSs and decoding strategy of users for code-domain  NOMA networks so as to minimize bit error rate \cite{kim2018deep}, where users upload the learned result to the corresponding BSs and the BSs forward their unified learned result to a server.
%Channel Estimation:
%message passing
%
%compressive sensing
%Signal Detection: multi-user detection, coding

\subsubsection{User behaviour prediction in NOMA}
%\item\textbf{Distributed learning for User Behaviour Prediction in NOMA (3 pages)}
%\begin{itemize}
%\item  User mobility prediction for NOMA with federated learning.
%\item  User distribution prediction for NOMA with federated learning.
%\item  Quality of service prediction with federated learning in the design of NOMA.
%\end{itemize}
Due to the heterogeneous quality-of-service requirement of users in NOMA, where users in the same cluster forming NOMA should have diversified channel gains and quality of service, user behaviour prediction is of great importance for the implementation of NOMA networks.
To predict the users behaviors such as mobility patterns, each user in FSL scheme performs a supervised learning algorithm to train the learning model using its own user behavior data and upload the trained model to the BS via NOMA. Then the BS generates and broadcasts the unified learning model parameters to all users by using NOMA. Based on the mobility patter predictions, the users can dynamically choose subchannel to upload data in the uplink, the BS dynamically allocates multiple subchannels to multiple users in the downlink, and multiple users which occupy the same subchannel can perform NOMA.
For multiple BSs to predict the quality of service of users \cite{Samarakoon2018Federated} in FSL, each BS uses supervised learning algorithm based on its stored information such as users' requested data, gender, job, and device type and all BSs transmit the learning model results to a server via NOMA to get a unified federated learning model.

\section{Research Directions and Challenges}

FL ensures that the resource allocation or behavior prediction problem can be solved in a distributed manner for wireless networks.
The utilization of FL for wireless networks has the following five main directions and challenges.

%% 1023
\begin{enumerate}
\item[1.] Convergence analysis: Due to the limited number of resource blocks (RBs) in a wireless
network, only a subset of users can be selected to transmit their local FL model parameters to the BS at
each learning step. Moreover, since each user has unique training data samples, the BS prefers to include
all local user FL models to generate a converged global FL model. Hence, the FL performance and
convergence time will be significantly affected by the user selection scheme.
Most of the FL convergence proof is established on the assumption that the loss function in convex \cite{chen2020convergence,yang2019scheduling}.
However, the loss function of the popular neural network is non-convex. It is a challenge to investigate the convergence rate for FL with non-convex loss function.
\item[2.] Privacy and security:
In FL, the raw data set at each user can be protected since only the local FL model is transmitted to the BS.
However, it is also possible for Eavesdropper to reconstruct the raw data approximately, especially when the local and global model parameters are not well protected \cite{ma2020safeguarding}.
Besides, the local FL model may leak private information.
In FL, the security can be classified into two categories: global security and local security. Global privacy requires that the model updates generated at each round are private to all untrusted third
parties other than the central server, while local privacy further requires that the updates are also private to
the server.
\item[3.] Asynchronous communication: Fl involves the information exchange between wireless devices and the BS. Synchronous communication methods are simple, which introduce stragglers among all devices.
    Asynchronous schemes are an attractive approach
to mitigate stragglers in heterogeneous environments.
server. While asynchronous parameter servers have been successful in distributed data centers, classical bounded-delay assumptions can be unrealistic in federated settings.
\item[4.] Non-iid device: Challenges arise when training federated models from data that is not identically distributed across devices,
both in terms of modeling the data, and in terms of analyzing the convergence
behavior of associated training procedures.
limited computation capacity at some wireless devices
causes delays.
\item[5.] Joint communication and computation design:  To deploy FL in wireless networks, devices are required to transmit their multimedia data or local training results over unreliable wireless links. This exposes the performance of learning and inference to degradation caused by limited radio resources (e.g., power, time and bandwidth). This makes it important to jointly manage communication and computation resources for efficient and robust FL.
\end{enumerate}

\section{Open Problems and Future Directions}

This section is to discuss open research
problems in each one of the covered areas, in order to shed
light on future opportunities. Despite a
considerable number of studies on FL, there
are still many key open problems that must be investigated about FL for wireless communications.

\begin{enumerate}
\item[1.] Convergence:  For FL convergence rate, there are still some key problems.
For example, there is a need for exact/more accurate convergence formulation with less assumptions and approximations \cite{chen2020convergence}, which should consistent with real FL experiment data.
Although there are some studies in this area, most of them related to convex loss function.
%, one single task, or a single cell.
Besides, due to the heterogeneous property of quality of service, it is possible to simultaneously conduct multi-task FL.
In addition,  for large-scale system, the multi-cell and multi-hop FL should be considered, which require one must have more insights on FL convergence analysis.
Moreover, one challenge is to study the mobility of wireless devices for FL convergence.
Due to the mobility, the channel gains between devices and BS are dynamically changed and it is possible that some devices will quit the FL process due to serious channel state information, which affects the convergence of the whole FL process.

\item[2.] Privacy and security: In terms of open problems for privacy and security, there is a need for the following study: privacy protection at each user, privacy protection at the BS, and security for the whole FL algorithm.
For privacy protection at each user and the BS, one of the key problem is to study the coding scheme and physical layer security technique.
For security of the whole FL algorithm, there is a need to study the encryption (such as quantum key distribution) and defender.
\item[3.] Performance evaluation:
One of the challenges is to investigate communication bandwidth for FL delay performance. FL on mobile phones relies on wireless communication to collaboratively learn a machine learning model. Although compute resources of mobile phones are becoming increasingly powerful, the bandwidth of wireless communication has not increased as much. As such, the bottleneck is shifted from computation to communication. As a consequence, limited communication bandwidth could incur long communication latency, and thus could significantly slow down the convergence time of the FL process.

\item[4.] FL for emerging technologies: The interplay between FL and emerging technologies introduces new challenges.
For instance, the very high propagation attenuations in THz can affect the convergence analysis.
For instance, in satellite communication, FL can used to optimize beam and location of the satellite.
For brain-computer, one of the challenge is to use FL extract deep knowledge of the brain's neural network.
In quantum communication, there is a need to use FL optimize the parameters (such as base probability) for quantum key distribution.
\end{enumerate}

\section{Conclusions}

In this tutorial, we have provided a comprehensive study on
the use of FL for wireless networks. We have investigated
two main classifications of FL, namely, FRL and FSL.
Besides, we have provided the motivation applications of using FL for wireless communications.
Meanwhile, we have described the
techniques needed to meet the challenges of using FL for wireless communications. Such an in-depth
study on FL for wireless communications provides unique
guidelines for optimizing, designing and operating FL-based
wireless communication systems.

\bibliographystyle{IEEEtran}
\bibliography{IEEEfull,Reference}

\begin{thebibliography}{10}
\providecommand{\url}[1]{#1}
\csname url@samestyle\endcsname
\providecommand{\newblock}{\relax}
\providecommand{\bibinfo}[2]{#2}
\providecommand{\BIBentrySTDinterwordspacing}{\spaceskip=0pt\relax}
\providecommand{\BIBentryALTinterwordstretchfactor}{4}
\providecommand{\BIBentryALTinterwordspacing}{\spaceskip=\fontdimen2\font plus
\BIBentryALTinterwordstretchfactor\fontdimen3\font minus
  \fontdimen4\font\relax}
\providecommand{\BIBforeignlanguage}[2]{{%
\expandafter\ifx\csname l@#1\endcsname\relax
\typeout{** WARNING: IEEEtran.bst: No hyphenation pattern has been}%
\typeout{** loaded for the language `#1'. Using the pattern for}%
\typeout{** the default language instead.}%
\else
\language=\csname l@#1\endcsname
\fi
#2}}
\providecommand{\BIBdecl}{\relax}
\BIBdecl

\bibitem{saad2019vision}
W.~Saad, M.~Bennis, and M.~Chen, ``A vision of {6G} wireless systems:
  {A}pplications, trends, technologies, and open research problems,''
  \emph{IEEE Network}, vol.~34, no.~3, pp. 134--142, 2020.

\bibitem{chen2019joint}
M.~Chen, Z.~Yang, W.~Saad, C.~Yin, H.~V. Poor, and S.~Cui, ``A joint learning
  and communications framework for federated learning over wireless networks,''
  \emph{IEEE Trans. Wireless Commun.}, 2020, to appear.

\bibitem{konevcny2016federated}
J.~Kone{\v{c}}n{\`y}, H.~B. McMahan, D.~Ramage, and P.~Richt{\'a}rik,
  ``Federated optimization: {D}istributed machine learning for on-device
  intelligence,'' \emph{arXiv preprint arXiv:1610.02527}, 2016.

\bibitem{huang2020federated}
M.~Bennis, M.~Debbah, K.~Huang, and Z.~Yang, ``Communication technologies for
  efficient edge learning,'' \emph{IEEE Commun. Magazine}, 2020 (To appear).

\bibitem{zhu2019broadband}
G.~Zhu, Y.~Wang, and K.~Huang, ``Broadband analog aggregation for low-latency
  federated edge learning,'' \emph{IEEE Trans. Wireless Commun.}, vol.~19,
  no.~1, pp. 491--506, 2019.

\bibitem{zhu2020one}
G.~Zhu, Y.~Du, D.~Gunduz, and K.~Huang, ``One-bit over-the-air aggregation for
  communication-efficient federated edge learning: {D}esign and convergence
  analysis,'' \emph{arXiv preprint arXiv:2001.05713}, 2020.

\bibitem{zeng2020energy}
Q.~Zeng, Y.~Du, K.~Huang, and K.~K. Leung, ``Energy-efficient resource
  management for federated edge learning with {CPU-GPU} heterogeneous
  computing,'' \emph{arXiv preprint arXiv:2007.07122}, 2020.

\bibitem{amiri2020machine}
M.~M. Amiri and D.~G{\"u}nd{\"u}z, ``Machine learning at the wireless edge:
  Distributed stochastic gradient descent over-the-air,'' \emph{IEEE Trans.
  Signal Process.}, vol.~68, pp. 2155--2169, 2020.

\bibitem{gunduz2020communicate}
D.~Gunduz, D.~B. Kurka, M.~Jankowski, M.~M. Amiri, E.~Ozfatura, and
  S.~Sreekumar, ``Communicate to learn at the edge,'' \emph{arXiv preprint
  arXiv:2009.13269}, 2020.

\bibitem{amiri2020federated}
M.~M. Amiri and D.~G{\"u}nd{\"u}z, ``Federated learning over wireless fading
  channels,'' \emph{IEEE Trans. Wireless Commun.}, vol.~19, no.~5, pp.
  3546--3557, 2020.

\bibitem{hosseinalipour2020federated}
S.~Hosseinalipour, C.~G. Brinton, V.~Aggarwal, H.~Dai, and M.~Chiang, ``From
  federated learning to fog learning: Towards large-scale distributed machine
  learning in heterogeneous wireless networks,'' \emph{arXiv preprint
  arXiv:2006.03594}, 2020.

\bibitem{hosseinalipour2020multi}
S.~Hosseinalipour, S.~S. Azam, C.~G. Brinton, N.~Michelusi, V.~Aggarwal, D.~J.
  Love, and H.~Dai, ``Multi-stage hybrid federated learning over large-scale
  wireless fog networks,'' \emph{arXiv preprint arXiv:2007.09511}, 2020.

\bibitem{jin2020design}
R.~Jin, X.~He, and H.~Dai, ``On the design of communication efficient federated
  learning over wireless networks,'' \emph{arXiv preprint arXiv:2004.07351},
  2020.

\bibitem{liu2020privacy}
D.~Liu and O.~Simeone, ``Privacy for free: Wireless federated learning via
  uncoded transmission with adaptive power control,'' \emph{arXiv preprint
  arXiv:2006.05459}, 2020.

\bibitem{kassab2020federated}
R.~Kassab and O.~Simeone, ``Federated generalized bayesian learning via
  distributed stein variational gradient descent,'' \emph{arXiv preprint
  arXiv:2009.06419}, 2020.

\bibitem{kairouz2019advances}
P.~Kairouz, H.~B. McMahan, B.~Avent, A.~Bellet, M.~Bennis, A.~N. Bhagoji,
  K.~Bonawitz, Z.~Charles, G.~Cormode, R.~Cummings \emph{et~al.}, ``Advances
  and open problems in federated learning,'' \emph{arXiv preprint
  arXiv:1912.04977}, 2019.

\bibitem{samarakoon2019distributed}
S.~Samarakoon, M.~Bennis, W.~Saad, and M.~Debbah, ``Distributed federated
  learning for ultra-reliable low-latency vehicular communications,''
  \emph{IEEE Transactions on Communications}, vol.~68, no.~2, pp. 1146--1159,
  2019.

\bibitem{yang2019eeFL}
Z.~Yang, M.~Chen, W.~Saad, C.~S. Hong, and M.~Shikh-Bahaei, ``Energy efficient
  federated learning over wireless communication networks,'' \emph{IEEE Trans.
  Wireless Commun.}, 2020 (To appear).

\bibitem{liu2019lifelong}
B.~Liu, L.~Wang, M.~Liu, and C.~Xu, ``Lifelong federated reinforcement
  learning: a learning architecture for navigation in cloud robotic systems,''
  \emph{arXiv preprint arXiv:1901.06455}, 2019.

\bibitem{li2020federated}
T.~Li, A.~K. Sahu, A.~Talwalkar, and V.~Smith, ``Federated learning:
  Challenges, methods, and future directions,'' \emph{IEEE Signal Process.
  Magazine}, vol.~37, no.~3, pp. 50--60, 2020.

\bibitem{lim2020federated}
W.~Y.~B. Lim, N.~C. Luong, D.~T. Hoang, Y.~Jiao, Y.-C. Liang, Q.~Yang,
  D.~Niyato, and C.~Miao, ``Federated learning in mobile edge networks: {A}
  comprehensive survey,'' \emph{IEEE Communications Surveys \& Tutorials},
  2020.

\bibitem{murshed2019machine}
M.~Murshed, C.~Murphy, D.~Hou, N.~Khan, G.~Ananthanarayanan, and F.~Hussain,
  ``Machine learning at the network edge: {A} survey,'' \emph{arXiv preprint
  arXiv:1908.00080}, 2019.

\bibitem{wang2019edge}
X.~Wang, Y.~Han, C.~Wang, Q.~Zhao, X.~Chen, and M.~Chen, ``In-edge {AI}:
  Intelligentizing mobile edge computing, caching and communication by
  federated learning,'' \emph{IEEE Network}, vol.~33, no.~5, pp. 156--165,
  2019.

\bibitem{park2019wireless}
J.~Park, S.~Samarakoon, M.~Bennis, and M.~Debbah, ``Wireless network
  intelligence at the edge,'' \emph{Proceedings of the IEEE}, vol. 107, no.~11,
  pp. 2204--2239, 2019.

\bibitem{aledhari2020federated}
M.~Aledhari, R.~Razzak, R.~M. Parizi, and F.~Saeed, ``Federated learning: A
  survey on enabling technologies, protocols, and applications,'' \emph{IEEE
  Access}, vol.~8, pp. 140\,699--140\,725, 2020.

\bibitem{8952884}
K.~{Yang}, T.~{Jiang}, Y.~{Shi}, and Z.~{Ding}, ``Federated learning via
  over-the-air computation,'' \emph{IEEE Trans. Wireless Commun.}, pp. 1--1,
  2020.

\bibitem{basar2019wireless}
E.~Basar, M.~Di~Renzo, J.~de~Rosny, M.~Debbah, M.-S. Alouini, and R.~Zhang,
  ``Wireless communications through reconfigurable intelligent surfaces,''
  \emph{arXiv preprint arXiv:1906.09490}, 2019.

\bibitem{zhang2019capacity}
S.~Zhang and R.~Zhang, ``Capacity characterization for intelligent reflecting
  surface aided {MIMO} communication,'' \emph{arXiv preprint arXiv:1910.01573},
  2019.

\bibitem{8580675}
S.~{Hu}, K.~{Chitti}, F.~{Rusek}, and O.~{Edfors}, ``User assignment with
  distributed large intelligent surface ({LIS}) systems,'' in \emph{Proc. IEEE
  Int. Symposium Personal, Indoor Mobile Radio Commun.}, Bologna, Italy, Sep.
  2018, pp. 1--6.

\bibitem{pan2019intelligent2}
C.~Pan, H.~Ren, K.~Wang, W.~Xu, M.~Elkashlan, A.~Nallanathan, and L.~Hanzo,
  ``Intelligent reflecting surface for multicell {MIMO} communications,''
  \emph{arXiv preprint arXiv:1907.10864}, 2019.

\bibitem{nadeem2019large}
Q.-U.-A. Nadeem, A.~Kammoun, A.~Chaaban, M.~Debbah, and M.-S. Alouini, ``Large
  intelligent surface assisted {MIMO} communications,'' \emph{arXiv preprint
  arXiv:1903.08127}, 2019.

\bibitem{wei2020joint}
L.~Wei, C.~Huang, G.~C. Alexandropoulos, Z.~Yang, C.~Yuen, and Z.~Zhang,
  ``Joint channel estimation and signal recovery in {RIS}-assisted multi-user
  {MISO} communications,'' \emph{arXiv preprint arXiv:2011.13116}, 2020.

\bibitem{8741198}
C.~{Huang}, A.~{Zappone}, G.~C. {Alexandropoulos}, M.~{Debbah}, and C.~{Yuen},
  ``Reconfigurable intelligent surfaces for energy efficiency in wireless
  communication,'' \emph{IEEE Trans. Wireless Commun.}, vol.~18, no.~8, pp.
  4157--4170, Aug. 2019.

\bibitem{chongwenDL2020}
C.~{Huang}, R.~{Mo}, and C.~{Yuen}, ``Reconfigurable intelligent surface
  assisted multiuser {{MISO}} systems exploiting deep reinforcement learning,''
  \emph{IEEE J. Sel. Areas Commun.}, vol.~38, pp. 1--1, 2020.

\bibitem{huang2019holographic}
C.~Huang, S.~Hu, G.~C. Alexandropoulos, A.~Zappone, C.~Yuen, R.~Zhang,
  M.~Di~Renzo, and M.~Debbah, ``Holographic {MIMO} surfaces for {6G} wireless
  networks: {O}pportunities, challenges, and trends,'' \emph{arXiv preprint
  arXiv:1911.12296}, 2019.

\bibitem{yu2019robust}
X.~Yu, D.~Xu, Y.~Sun, D.~W.~K. Ng, and R.~Schober, ``Robust and secure wireless
  communications via intelligent reflecting surfaces,'' \emph{arXiv preprint
  arXiv:1912.01497}, 2019.

\bibitem{zheng2020double}
B.~Zheng, C.~You, and R.~Zhang, ``Double-irs assisted multi-user mimo:
  Cooperative passive beamforming design,'' \emph{arXiv preprint
  arXiv:2008.13701}, 2020.

\bibitem{chaccour2020risk}
C.~Chaccour, M.~N. Soorki, W.~Saad, M.~Bennis, and P.~Popovski, ``Risk-based
  optimization of virtual reality over terahertz reconfigurable intelligent
  surfaces,'' \emph{arXiv preprint arXiv:2002.09052}, 2020.

\bibitem{hum2013reconfigurable}
S.~V. Hum and J.~Perruisseau-Carrier, ``Reconfigurable reflectarrays and array
  lenses for dynamic antenna beam control: {A} review,'' \emph{IEEE Trans.
  Antennas Prop.}, vol.~62, no.~1, pp. 183--198, Jan. 2013.

\bibitem{huang2014relay}
J.~Huang, Q.~Li, Q.~Zhang, G.~Zhang, and J.~Qin, ``Relay beamforming for
  amplify-and-forward multi-antenna relay networks with energy harvesting
  constraint,'' \emph{IEEE Signal Process. Lett.}, vol.~21, no.~4, pp.
  454--458, Apr. 2014.

\bibitem{ntontin2019reconfigurable}
K.~Ntontin, M.~Di~Renzo, J.~Song, F.~Lazarakis, J.~de~Rosny, D.-T. Phan-Huy,
  O.~Simeone, R.~Zhang, M.~Debbah, G.~Lerosey, M.~Fink, S.~Tretyakov, and
  S.~Shamai, ``Reconfigurable intelligent surfaces vs. relaying: {D}ifferences,
  similarities, and performance comparison,'' \emph{arXiv preprint
  arXiv:1908.08747}, 2019.

\bibitem{elbir2020federated}
A.~M. Elbir and S.~Coleri, ``Federated learning for channel estimation in
  conventional and irs-assisted massive mimo,'' \emph{arXiv preprint
  arXiv:2008.10846}, 2020.

\bibitem{yang2020federated}
K.~Yang, Y.~Shi, Y.~Zhou, Z.~Yang, L.~Fu, and W.~Chen, ``Federated machine
  learning for intelligent iot via reconfigurable intelligent surface,''
  \emph{arXiv preprint arXiv:2004.05843}, 2020.

\bibitem{ni2020federated}
W.~Ni, Y.~Liu, Z.~Yang, H.~Tian, and X.~Shen, ``Federated learning in multi-ris
  aided systems,'' \emph{arXiv preprint arXiv:2010.13333}, 2020.

\bibitem{huang2019indoor}
C.~Huang, G.~C. Alexandropoulos, C.~Yuen, and M.~Debbah, ``Indoor signal
  focusing with deep learning designed reconfigurable intelligent surfaces,''
  in \emph{2019 IEEE 20th International Workshop on Signal Processing Advances
  in Wireless Communications (SPAWC)}.\hskip 1em plus 0.5em minus 0.4em\relax
  IEEE, 2019, pp. 1--5.

\bibitem{huang2020hybrid}
C.~Huang, Z.~Yang, G.~C. Alexandropoulos, K.~Xiong, L.~Wei, C.~Yuen, and
  Z.~Zhang, ``Hybrid beamforming for ris-empowered multi-hop terahertz
  communications: A drl-based method,'' \emph{arXiv preprint arXiv:2009.09380},
  2020.

\bibitem{xie2020deep}
H.~Xie, Z.~Qin, G.~Y. Li, and B.-H. Juang, ``Deep learning enabled semantic
  communication systems,'' \emph{arXiv preprint arXiv:2006.10685}, 2020.

\bibitem{xie2020lite}
H.~Xie and Z.~Qin, ``A lite distributed semantic communication system for
  internet of things,'' \emph{arXiv preprint arXiv:2007.11095}, 2020.

\bibitem{chen2019federated}
M.~Chen, O.~Semiari, W.~Saad, X.~Liu, and C.~Yin, ``Federated echo state
  learning for minimizing breaks in presence in wireless virtual reality
  networks,'' \emph{IEEE Transactions on Wireless Communications}, vol.~19,
  no.~1, pp. 177--191, 2019.

\bibitem{8755300}
M.~{Chen}, U.~{Challita}, W.~{Saad}, C.~{Yin}, and M.~{Debbah}, ``Artificial
  neural networks-based machine learning for wireless networks: {A} tutorial,''
  \emph{IEEE Commun. Surveys Tutorials}, vol.~21, no.~4, pp. 3039--3071,
  Fourthquarter 2019.

\bibitem{Zhiguo2015Mag}
Z.~Ding, Y.~Liu, J.~Choi, Q.~Sun, M.~Elkashlan, C.-L. I, and H.~V. Poor,
  ``Application of non-orthogonal multiple access in {LTE} and {5G} networks,''
  \emph{{IEEE} Commun. Mag.}, vol.~55, no.~2, pp. 185--191, Feb. 2017.

\bibitem{Liu2017Proceeding}
Y.~{Liu}, Z.~{Qin}, M.~{Elkashlan}, Z.~{Ding}, A.~{Nallanathan}, and
  L.~{Hanzo}, ``Nonorthogonal multiple access for {5G} and beyond,''
  \emph{Proceedings of the IEEE}, vol. 105, no.~12, pp. 2347--2381, Dec. 2017.

\bibitem{Liu2018MIMONOMA}
Y.~{Liu}, H.~{Xing}, C.~{Pan}, A.~{Nallanathan}, M.~{Elkashlan}, and
  L.~{Hanzo}, ``Multiple-antenna-assisted non-orthogonal multiple access,''
  \emph{{IEEE} Wireless Commun.}, vol.~25, no.~2, pp. 17--23, Apr. 2018.

\bibitem{Qin2018Mag}
Z.~{Qin}, X.~{Yue}, Y.~{Liu}, Z.~{Ding}, and A.~{Nallanathan}, ``User
  association and resource allocation in unified {NOMA} enabled heterogeneous
  ultra dense networks,'' \emph{{IEEE} Commun. Mag.}, vol.~56, no.~6, pp.
  86--92, June 2018.

\bibitem{yang2017on}
Z.~Yang, W.~Xu, C.~Pan, Y.~Pan, and M.~Chen, ``On the optimality of power
  allocation for {NOMA} downlinks with individual {QoS} constraints,''
  \emph{IEEE Commun. Lett.}, vol.~21, no.~7, pp. 1649--1652, July 2017.

\bibitem{Yang2018Power}
Z.~Yang, C.~Pan, W.~Xu, Y.~Pan, M.~Chen, and M.~Elkashlan, ``Power control for
  multi-cell networks with non-orthogonal multiple access,'' \emph{IEEE Trans.
  Wireless Commun.}, vol.~17, no.~2, pp. 927--942, Feb. 2018.

\bibitem{Shin2017Mag}
W.~{Shin}, M.~{Vaezi}, B.~{Lee}, D.~J. {Love}, J.~{Lee}, and H.~V. {Poor},
  ``Non-orthogonal multiple access in multi-cell networks: Theory, performance,
  and practical challenges,'' \emph{{IEEE} Commun. Mag.}, vol.~55, no.~10, pp.
  176--183, Oct. 2017.

\bibitem{ni2020resource}
W.~Ni, X.~Liu, Y.~Liu, H.~Tian, and Y.~Chen, ``Resource allocation for
  multi-cell irs-aided noma networks,'' \emph{arXiv preprint arXiv:2006.11811},
  2020.

\bibitem{andrieu2003introduction}
C.~Andrieu, N.~De~Freitas, A.~Doucet, and M.~I. Jordan, ``An introduction to
  {MCMC} for machine learning,'' \emph{Machine Learning}, vol.~50, no. 1-2, pp.
  5--43, Jan. 2003.

\bibitem{freeman2000learning}
W.~T. Freeman, E.~C. Pasztor, and O.~T. Carmichael, ``Learning low-level
  vision,'' \emph{International Journal of Computer Vision}, vol.~40, no.~1,
  pp. 25--47, Oct. 2000.

\bibitem{collobert2008unified}
R.~Collobert and J.~Weston, ``A unified architecture for natural language
  processing: Deep neural networks with multitask learning,'' in \emph{Proc. of
  the International Conference on Machine Learning}, New York, NY, USA, July
  2008, pp. 160--167.

\bibitem{bishop2006pattern}
C.~M. Bishop, \emph{Pattern recognition and machine learning}.\hskip 1em plus
  0.5em minus 0.4em\relax Springer, 2006.

\bibitem{7001098}
H.~Lee, M.~Wicke, B.~Kusy, O.~Gnawali, and L.~Guibas, ``Predictive data
  delivery to mobile users through mobility learning in wireless sensor
  networks,'' \emph{IEEE Trans. Veh. Technol.}, vol.~64, no.~12, pp.
  5831--5849, Dec 2015.

\bibitem{8219722}
L.~Yao, A.~Chen, J.~Deng, J.~Wang, and G.~Wu, ``A cooperative caching scheme
  based on mobility prediction in vehicular content centric networks,''
  \emph{IEEE Trans. Veh. Technol.}, vol.~67, no.~6, pp. 5435--5444, June 2018.

\bibitem{chen2016caching}
M.~Chen, M.~Mozaffari, W.~Saad, C.~Yin, M.~Debbah, and C.~S. Hong, ``Caching in
  the sky: {P}roactive deployment of cache-enabled unmanned aerial vehicles for
  optimized quality-of-experience,'' \emph{IEEE J. Sel. Areas Commun.},
  vol.~35, no.~5, pp. 1046--1061, May 2017.

\bibitem{8372711}
J.~Yin, L.~Li, H.~Zhang, X.~Li, A.~Gao, and Z.~Han, ``A prediction-based
  coordination caching scheme for content centric networking,'' in \emph{Proc.
  of Wireless and Optical Communication Conference}, Hualien, Taiwan, April
  2018.

\bibitem{Yang2017Energy}
Z.~Yang, W.~Xu, H.~Xu, J.~Shi, and M.~Chen, ``Energy efficient non-orthogonal
  multiple access for machine-to-machine communications,'' \emph{IEEE Commun.
  Lett.}, vol.~21, no.~4, pp. 817--820, Apr. 2017.

\bibitem{Yang2018EEIoT}
Z.~Yang, W.~Xu, Y.~Pan, C.~Pan, and M.~Chen, ``Energy efficient resource
  allocation in machine-to-machine communications with multiple access and
  energy harvesting for {IoT},'' \emph{IEEE Internet Things J.}, vol.~5, no.~1,
  pp. 229--245, Feb. 2018.

\bibitem{Cui2018Unspuer}
J.~{Cui}, Z.~{Ding}, P.~{Fan}, and N.~{Al-Dhahir}, ``Unsupervised machine
  learning-based user clustering in millimeter-wave-{NOMA} systems,''
  \emph{IEEE Trans. Wireless Commun.}, vol.~17, no.~11, pp. 7425--7440, Nov.
  2018.

\bibitem{li2013parameter}
M.~Li, L.~Zhou, Z.~Yang, A.~Li, F.~Xia, D.~G. Andersen, and A.~Smola,
  ``Parameter server for distributed machine learning,'' in \emph{Big Learning
  {NIPS} Workshop}, vol.~6, 2013, p.~2.

\bibitem{bekkerman2011scaling}
R.~Bekkerman, M.~Bilenko, and J.~Langford, \emph{Scaling up machine learning:
  {P}arallel and distributed approaches}.\hskip 1em plus 0.5em minus
  0.4em\relax Cambridge University Press, 2011.

\bibitem{kim2018deep}
M.~Kim, N.-I. Kim, W.~Lee, and D.-H. Cho, ``Deep learning-aided {SCMA},''
  \emph{{IEEE} Commun. Lett.}, vol.~22, no.~4, pp. 720--723, 2018.

\bibitem{Samarakoon2018Federated}
S.~{Samarakoon}, M.~{Bennis}, W.~{Saad}, and M.~{Debbah}, ``Federated learning
  for ultra-reliable low-latency {V2V} communications,'' in \emph{Proc. IEEE
  Global Commun. Conf.}, Abu Dhabi, United Arab Emirates, Dec. 2018, pp. 1--7.

\bibitem{chen2020convergence}
M.~Chen, H.~V. Poor, W.~Saad, and S.~Cui, ``Convergence time optimization for
  federated learning over wireless networks,'' \emph{IEEE Trans. Wireless
  Commun.}, 2020 (To appear).

\bibitem{yang2019scheduling}
H.~H. Yang, Z.~Liu, T.~Q. Quek, and H.~V. Poor, ``Scheduling policies for
  federated learning in wireless networks,'' \emph{IEEE Transactions on
  Communications}, vol.~68, no.~1, pp. 317--333, 2019.

\bibitem{ma2020safeguarding}
C.~Ma, J.~Li, M.~Ding, H.~H. Yang, F.~Shu, T.~Q. Quek, and H.~V. Poor, ``On
  safeguarding privacy and security in the framework of federated learning,''
  \emph{IEEE Network}, 2020.

\end{thebibliography}
%\newpage

\end{document}